\documentclass [12pt,a4paper]{article}
\usepackage{graphicx}
\usepackage{psfrag}
\usepackage{graphics}
\usepackage{bm}
\usepackage{color}
\usepackage{verbatim,color,ulem}

\input amssym

\newcommand{\gsim}{\raise.3ex\hbox{$>$\kern-.75em\lower1ex\hbox{$\sim$}}}
\newcommand{\lsim}{\raise.3ex\hbox{$<$\kern-.75em\lower1ex\hbox{$\sim$}}}
\begin{document}

%\renewcommand{\thefootnote}{\fnsymbol{footnote}}

%-------------------------------------------------

% *********************************************
\begin{center}
{\Large\textbf{An extensive universe avoids phantom dark energy }}
\\[\baselineskip]
 {\large {C.C.\ Wong}\textsuperscript{1}},
 {\large {R.J.\ Rivers}\textsuperscript{2}}
 \\[0.5\baselineskip]
 1. Department of Electrical Engineering,
     City University of Hong Kong. H.K.
 \\
 2. Abdus Salam Centre for Theoretical Physics, Physics Department,
     Imperial College London, SW7 2AZ, U.K.
\\[0.7\baselineskip]
Corresponding E-mail: chicwong@cityu.edu.hk
%\\
%	Contributing author: chicwong@cityu.edu.hk.
\end{center}

\date{\today}
\begin{abstract}
\noindent A major challenge in cosmology and astrophysics is the phantom dark energy used to explain the data of DESI DR2 BAO. Assuming that the degrees of freedom of the universe are ($d=3$) extensive, we find that the linearised Hubble parameter in  Tsallis Entropic cosmology is consistent with Cosmic Chronometers (CC) and DESI DR2 data up to $z\leq 2.33$, avoiding the need for phantom dark energy while keeping the observed CMB acoustic angular scale constraint. It also obviates the need for a significant density of CDM.
\end{abstract}

%\pacs{??}

%\maketitle

\section{Introduction}
In this paper we shall attempt an explanation of a major challenge in cosmology and astrophysics; the 'phantom' dark energy invoked to explain the DESI DR2 data \cite{karim}-\cite{scherer}. Even though it is modelled to occur above $z>0.5$,  if taken seriously, phantom dark energy has negative kinetic energy which in principle could have led to the 'ripping' of the  universe \cite{caldwell}.
The framework that we adopt is Tsallis Entropic cosmology, of which there is an increasing literature \cite{tsallis}-\cite{moha2},
which provides a generalisation of the Bekenstein-Hawking area-law black-hole entropy \cite{beken1}-\cite{hawking2} to  extensive degrees of freedom in $d=3$ dimensions. Entropic cosmology most simply generalises the Friedmann equations by modifying the stress-energy tensor, while leaving Einstein's equations unchanged.
\\\\
In this work, under the dynamics of this new paradigm we consider Hubble parameter values within the redshift range of DESI \cite{karim}.  In Section 2 we recall recent results in Tsallis Entropic cosmology in the literature and find a new solution for a multi-fluid universe. In Section 3 we calculate the Hubble parameter $H(z)$ in this model solution within the constraint of CMB observed angular scales and compare with observations. In the absence of anomalous dimensions \cite{barrow}, extensivity imposes a linearisation of the Hubble constant which is well realised empirically \cite{jizba}. This Hubble linearisation is consistent with Cosmic Chronometers (CC) and DESI DR2 data up to $z\leq 2.33$, while keeping the observed CMB acoustic angular scale constraint, thereby obviating the need for phantom dark energy. This is shown in Figs. 1 and 2.
In Section 4 we discuss something not yet mentioned, that the $\Lambda$CDM constraint
on the dark matter density parameter coming from the observed CMB acoustic
peak heights could be satisfied by our model without any fine-tuning.
Section 5 provides a summary.
\\\\
Before going into the equations we summarise the problems with the data in a little more detail. The  DESI DR2 study \cite{karim}-\cite{scherer} uses  baryonic acoustic oscillation (BAO) as a ruler at redshifts $0<z< 4$ based on a $\omega CDM$ cosmology. For CMB cosmological parameters and a variable dark energy density equation of state $\omega$, it is found \cite{karim} that $\omega$ crosses the phantom boundary into the $\omega<-1$ region at redshift $z > 0.5$ up to $z\sim 4$. For $\omega<-1$, the dark energy violates the null energy condition for an extended period in the cosmological history. Based on the late time Hubble parameter $H(z)=H_0\sqrt{ \Omega_m(1+z)^3+\Omega_{\Lambda}}$ where $\Omega_m,\:\Omega_{\Lambda}$ are the density parameters of total matter and dark energy from CMB observation, the observed $H(z)$ value at different redshifts leads to a Hubble constant $H_0 =0.68-0.74$ \cite{jia}-\cite{krishnan}. This extrapolation suggests that the observed $H(z)$ (at $z\leq 2.33$) is incompatible with the sum of terms $\Omega_m(1+z)^3+\Omega_{\Lambda} $  as constrained by the CMB data.
\\\\
To avoid DE phantom crossing at low $z$, either the matter sector (or its equation of state) or the dark energy sector will need to be modified.
We have adopted the former in keeping with our avoidance of CDM but Dynamical dark matter proposals have already been put forward \cite{loeb2}-\cite{dwang}. Sterile neutrino hot dark matter cosmology $\nu HDM$ has been investigated to give $h=0.56$ \cite{kroupa6}. $\Lambda_S CDM$ \cite{akarsu}  in which the dark energy changes sign at $z\sim 2$ has also been proposed, as are interactions between DM and DE \cite{zimdahl}. Other ideas may be found in \cite{lopez}.
\\\\
We do not need CDM in our analysis.

\section{Entropic cosmology}
The starting point is the Bekenstein-Hawking entropy $S_{BH}$ \cite{beken1}-\cite{hawking2} based on its statistical connection with Boltzmann-Gibbs (BG) entropy
\begin{equation}
S_{BH} \equiv S_{BG}\equiv k_BS_1 = k_B\sum_I^W p_i \ln (\frac{1}{ p_i}),
\end{equation}
where $S_1$ is Shannon entropy and $W$ is the total number of accessible microscopic states of a system. One obtains $ S_1=\ln W$ when every microstate is equally probable, such that $p_i =1/W$ for all $i$. Bekenstein finds that $S_{BG}$ is extensive in the black hole (BH) horizon area (but non-extensive in the associated volume). However, the non-extensivity of $S_{BH}$ when considered as a $d=3$ object and the fact that BG entropy is applicable to weak and short range interactions amongst the system elements has led to the study of non-additive but extensive entropies. For extensivity in a 3-d universe we require that
\begin{equation}
 S_{\delta} \asymp (\ln W)^{\delta}.
 \label{Wd}
\end{equation}
where $\delta = 3/2$.
\\\\
There are two straightforward ways to realise this. In Tsallis and Cirto (TC) \cite{tsallis}, based on \cite{ tsallis2}, Shannon entropy was first generalised as
\begin{equation}
S^{TC}_{\delta} = \sum_I^W p_i \ln (\frac{1}{ p_i})^{\delta}.
\label{sadelta}
\end{equation}
However,whereas $S^{TC}_{\delta}$ is extensive for $\delta = 3/2$ it is not composable  which precludes its consistent use within the framework of statistical mechanics \cite{jensen_jizba}.
\\\\
A simpler alternative is \cite{tsallis3,jensen_jizba}
\begin{equation}
 S_{\delta} = S_1^{\delta}.
\end{equation}
which is both extensive and composable for $\delta = 3/2$.
\\\\
For both choices we recover Eq.(\ref{Wd}), which is all we need for modified Friedmann equations. In either case, long range correlations amongst system elements are given as plausible reasons for this class of entropy.
\\\\
There is one further complication.The Barrow entropy toy model for black holes \cite{barrow}  which has no statistical roots permits
\begin{equation}
S_B\propto S_1^{1+\frac{\Delta}{2}}
\end{equation}
where $\Delta$ plays the role of an anomalous dimension (e.g for black hole surfaces a Hausdorff dimension of $2+ \Delta$ due to fractal behaviour). Inserting this microstate scaling into 3-d extensive entropy gives a final entropy
\begin{equation}
 S = S_1^{(1 + \Delta/2)\delta}
\end{equation}
where extensivity requires
\begin{equation}
(1 + \Delta/2)\delta = \frac{3}{2} \Leftrightarrow \delta = \frac{3}{2 + \Delta}
\end{equation}
Although not the most simple formalism, to conform  with the literature  \cite{jizba0}-\cite{moha2} in what follows, we rewrite the entropy as
\begin{equation}
S_{\delta} =\gamma(\delta) A^{\delta},\:\:\:\:\:
\label{sadelta2}
\end{equation}
where $\gamma=\gamma(\delta)$ is a $\delta$ dependent constant and $A = 4\pi R^2_H$ is the Horizon area. The Tsallis-Cirto value \cite{tsallis} is $\delta=\frac{3}{2},\,\,{\Delta = 0}$.
\subsection{Friedmann revisited}
A connection between gravity and thermodynamics is put forward by Jacobson \cite{jacobson}, Verlinde \cite{verlinde}, Padmanabhan \cite{pad} and Sheykhi \cite{sheykhi} such that the Einstein equations can be obtained by the first law of thermodynamics.
Given a FLRW metric, a perfect fluid stress-momentum tensor with uniform density $\rho$ and pressure $P$ and a universe with apparent horizon of radius $R_H =H^{-1}$ for zero spatial curvature, the first law of thermodynamics used in Sheykhi \cite{sheykhi}
%and Jizba-Lambiase \cite{jizba0}-\cite{jizba}
is
\begin{equation}
dE= -TdS +WdV,\:\:\:
\label{1}
\end{equation}
where
\begin{equation}
E=\rho V,\:\:\:V=\frac{4\pi}{3} R_H^3,\:\:\:  W=\frac{1}{2}(\rho-P),:\:T=\frac{|\kappa|}{2\pi}
\end{equation}
and $\kappa$ is the surface gravity see \cite{akbar}
\begin{equation}
\kappa=-\frac{1}{2\pi R_H}\bigg (1-\frac{\dot{R}_H}{2HR_H}\bigg)
\end{equation}
Eq,(\ref{1}) becomes
\begin{equation}
V d\rho+\frac{1}{2} (\rho+P)dV=-TdS
\label{2}
\end{equation}
Taking into account the continuity equation
\begin{equation}
\dot{\rho}=-3H(\rho+P),\:\:\:\sum_i \dot\rho=-3H\sum_i (\rho_i+P_i).
\end{equation}
where the $\rho_i$ are different fluid components in the cosmic background.
\begin{equation}
\frac{1}{2} (\rho+P )dV=\frac{1}{2}\frac{\dot{R_H}}{R_H}(\rho+P) 4\pi R_H^3 dt=-\frac{\dot{R}_H}{2HR_H}Vd\rho,
\end{equation}
Eq.(\ref{2}) now becomes
\begin{equation}
\bigg(1-\frac{1}{2H}\frac{\dot{R_H}}{R_H}\bigg) V d\rho = -TdS =-\frac{1}{2\pi R_H} \bigg(1-\frac{\dot{R_H}}{2HR_H}\bigg) dS
\label{tds}
\end{equation}
which leads to the base relation between $\rho$ and $H$,
\begin{equation}
Vd\rho=-\frac{1}{2\pi R_H} dS.
\label{vrho}
\end{equation}
Since $V$ and $S$ depend only on $R_H$, one can see that Eq.(\ref{vrho}) is a differential version of a modified Friedmann equation.
Using Tsallis entropy Eq.(\ref{sadelta2}) and integrating gives
\begin{equation}
 R_H^{2\delta-4}= (H^2)^{2-\delta}=\bigg(\frac{2\pi (2-\delta)}{3\gamma \delta} (4\pi)^{1-\delta} \bigg)\rho.
\label{rhoH1}
\end{equation}
Eq.(\ref{rhoH1}) is not yet the Friedmann equation, however if one takes $\delta=1$, and $4\gamma G=1$, Eq.(\ref{rhoH1}) becomes the Friedmann equation $H^2=\frac{8 \pi G}{3}\rho$.
\\\\
At this point, Tsallis and other entropy cosmologies encounter a major uncertainty. From direct observations, radiation and baryons are observable, whereas CDM and dark energy are invisible and the $\Lambda CDM$ model works very well in describing the  CMB power spectrum. In this revised cosmology either CDM or dark energy needs to be modified. This is exactly the same problem encountered by modelling of $H(z)$ at $z<2.23$ to match data. Most entropy cosmologies choose dark energy as the source for modification   to explain the DESI DR2 data. We shall show below that in Tsallis-Cirto cosmology (TC) \cite{tsallis}, it is the CDM component that needs modification.
\\\\
Now we come to the main characteristic of entropy cosmology. For two subsystems $M$ and $N$ with the same $R_H$ (with the universe horizon surface area $ A=4\pi R_H^2$) as the combined $M+N$ system, the entropy of the combined system is
\begin{equation}
S_{M+N}=\gamma(\delta) A^{\delta} \neq S_M+S_N.
\end{equation}
Jizba-Lambiase (JL) \cite{jizba0}-\cite{jizba} take a different approach to investigate the Tsallis-Cirto cosmology. JL note that the most important difference between $S_{BG}$ and $S_{\delta}$ is that $S_{\delta}$ is not additive but  obeys the pseudo-additivity rule. That is, for two independent subsystems $M$ and $N$
\begin{equation}
S_{M+N,\delta}=\bigg( S_{M, \delta}^{1/\delta}+S_{N,\delta}^{1/\delta}\bigg) ^{\delta}.
\label{SAB}
\end{equation} Differentiating Eq.(\ref{SAB})  gives
\begin{equation}
d S_{M+N,\delta}^{1/\delta}= dS_{M,\delta}^{1/\delta} + dS_{N,\delta}^{1/\delta}.
\label{dS}
\end{equation}
or
\begin{equation}
S_{M+N, \delta}^{1/\delta-1} dS_{M+N, \delta}= S_{M, \delta}^{1/\delta-1} dS_{M, \delta}+S_{N, \delta}^{1/\delta -1} dS_{N, \delta}
\label{pseudoS}
\end{equation}
Caratheodory's theorem ensures that the total differential of entropy  $dS$ and the heat 1-form $\delta Q$ has the relation
\begin{equation}
dS_{\delta}=\mu \delta Q,\:\:\:
\end{equation}
where $\mu$ is an integrating factor. JL \cite{jizba} argue that the integrating factor takes the form
\begin{equation}
\mu=T^{-1} S_{\delta}^{1- 1/\delta}
\end{equation}
\begin{equation}
\delta Q=\frac{1}{\mu} dS_{\delta}=TS^{1/\delta-1}dS_{\delta}=TdS_{\delta}^{1/\delta}.
\label{jizS}
\end{equation}
Taking $TdS_{\delta}^{1/\delta}$ into Eq.(\ref{1}), the modified Friedmann equation takes  the form
\begin{equation}
(H^2)^{2-\frac{\alpha}{2}}=\frac{8\pi G^{\alpha/2} }{3} \bigg[G^{2-\alpha}\bigg(\frac{8\pi }{3} \rho\bigg)^{1-\frac{\alpha}{2}}\bigg]\rho=\frac{8\pi G^{\alpha/2} }{3}\rho_{\alpha}.
\label{otc11}
\end{equation}
where it has been convenient \cite{jizba} to introduce $\alpha=\frac{3}{\delta}$. We note that in Eq.(\ref{otc11}) we have $H^2 \propto \rho$ for $\delta = 3/2,\,\,\alpha =2$.
\\\\
In the radiation dominant epoch, we can rewrite the temperature of $H$ in the form \cite{jizba0}-\cite{jizba}
\begin{equation}
H=Q(T) H_{St. Cosm} (T), \:\:\: \rho =\frac{\pi^2 106}{30} T^4.
\end{equation}
where $T$ is the system temperature and $H_{St. Cosm}$ is the Hubble parameter of standard cosmology ($\Lambda CDM$) in the radiation dominant epoch and
\begin{equation}
Q(T) \propto \bigg(\frac{T}{M_p}\bigg)^{\nu}, \:\: \nu=\frac{2(\alpha-2)}{(4-\alpha)}.
\end{equation}
Here $M_p$ is the Planck mass and $Q(T)$ can be regarded as an amplification factor. In JL \cite{jizba}, using BBN data (therefore in the radiation dominant epoch where CDM is negligible) the observations support the Tsallis-Cirto cosmology with $\alpha\simeq 2$. That is, $\delta=\frac{3}{\alpha} \simeq \frac{3}{2}$, $\Delta \simeq 0$ and we take $\alpha =2$, $\delta = \frac{3}{2}$, $\Delta = 0$ henceforth.
Specifically, JL \cite{jizba} find
\begin{equation}
\,\,\,\,\,\,\,1.4957 \lesssim \delta \lesssim 1.4990 \,\,\Leftrightarrow\,\, 0.0013 \lesssim \Delta \lesssim 0.0057
\nonumber
\end{equation}
from the freeze-out temperature. Constraints on $\delta$ from primordial abundance of light elements gives comparable bounds e.g. for $^4He$ they find
\begin{equation}
^4He\,\,\,\,\,\,1.4989 \lesssim \delta \lesssim 1.5024 \,\,\Leftrightarrow\,\, -0.0033 \lesssim \Delta \lesssim 0.0014.
\nonumber
\end{equation}
\noindent See \cite{jizba} for more details.  That is, the data supports the constraint $\delta = 3/2$, whereby ($\alpha =2$).
\\\\
The question is therefore how the modification due to $\delta$ enters the Hubble parameter in a multiple fluid background.
\subsection{ Our solution at $\delta=3/2$}
Again, suppose that the subsystems $M$ and $N$ are made up of distinct fluids but having equal volume $V$ and radius $R_H$. In this case the subsystem entropies $S_{i,\delta}=(\gamma A^{\delta})_i=\gamma (4\pi R_H^2)_i^{\delta},\: (i=M, N)$,  are distinguished only by the content in the subsystem $i$. We notice that for the special case $\delta=3/2$,  the component (considered as entropic by JL) in front of $dS$ in Eq.(\ref{pseudoS})
\begin{equation}
S_{i, \delta}^{1/\delta-1}=(\gamma A^{\delta})_i^{1/\delta-1}=\gamma^{1/\delta-1}  (4\pi R_H^2)_i^{1-\delta}=\bigg(\frac{1}{2}\gamma^{-1/3} (4\pi)^{1/2}\bigg) \bigg(\frac{1}{2\pi R_H}\bigg)_i.
\end{equation}
We rewrite
\begin{equation}
\bigg(\frac{1}{2}\gamma^{-1/3} (4\pi)^{1/2}\bigg) \gamma =\Gamma
\end{equation}
such that $S_{i, \delta}=(\Gamma A^{\delta})_i$, Eq.(\ref{pseudoS}) then takes the form
\begin{equation}
\frac{1}{2\pi R_H} dS_{M+N, \delta}=\frac{1}{2\pi R_H} dS_{M, \delta}+\frac{1}{2\pi R_H} dS_{N, \delta}
\label{STC}
\end{equation}
We recall that $1/(2\pi R_H)$ from Eq.(16) plays the role of $T$, the temperature at the horizon, which means that the pseudo-additivity rule Eq.(\ref{SAB}) becomes additive for entropy differentials $dS$, under the {\it special} condition given in Tsallis-Cirto entropy $S=\gamma A^{3/2}$. Eq.(\ref{STC}) allows the (differential) modifed Friedmann equation Eq.(\ref{vrho}) to be written as
\begin{equation}
 Vd(\sum_i \rho_i)=-\frac{1}{2\pi R_H}dS=-\frac{1}{2\pi R_H}\sum_i dS_{i, \delta}
\label{HT}
\end{equation}
which is the sum of modified Friedmann equations for different single fluids,
\begin{equation}
Vd\rho_i=-\frac{1}{2\pi R_H } dS_{i, \delta}
\label{HTi}
\end{equation}
The solutions for Eq.(\ref{HTi}) take the form
\begin{equation}
(H_i^2)^{2 -\delta} =\alpha_i\bigg(\frac{8 \pi G}{3} \rho_i\bigg)^{2 - \delta}.
\label{Hi0}
\end{equation}
The authors of \cite{sheykhi} leave the $\alpha_i$ unspecified, whereas the authors of \cite{moha1} propose  that
\begin{equation}
\alpha_i =\Omega_{i,0}^{1-(2-\delta)},
\label{AO}
\end{equation}
where $\Omega_{i,0}$ is the density parameter of the $i$ component. This is a consistent solution assuming, for any single fluid, that the underlying statistics is not Boltzmann-Gibbs.
%The reason is the following: 
In \cite{moha1} it is argued that,  when $\delta_N$ is defined by $2-\delta=1/\delta_N$, ($\delta=3/2, \: \delta_N=2$), the continuity equation becomes
\begin{equation}
\rho_i=\rho_{i,0}a^{\frac{3}{\delta_N}(1+\omega_i)}
\end{equation}
Eq.(\ref{rhoH1})  takes the form
\begin{equation}
(H^2)^{1/\delta_N}=\bigg(\frac{8\pi G\rho_c}{3}\bigg)^{1/\delta_N} \beta \sum_i \rho_{i,0}^{1-1/\delta_N} \bigg(\rho_{i,0} a^{3(1+\omega_i)}\bigg)^{1/\delta_N}
\label{moha}
\end{equation}
On dimensional grounds the proposed solution \cite{moha1} is $\beta^{-1}=\rho_c^{1-1/\delta_N}$ so that Eq.(\ref{moha}) becomes
\begin{equation}
(H^2)^{1/\delta_N}=\bigg(\frac{8 \pi G\rho_c}{3}\bigg)^{1/\delta_N}\sum_i \Omega_{i,0}^{1-1/\delta_N} \bigg(\rho_{i,0} a^{3(1+\omega_i)}\bigg)^{1/\delta_N}=\sum_i \Omega_{i,0}^{1-1/\delta_N} \rho_i^{1/\delta_N}
\label{moha1}
\end{equation}
\begin{equation}
H =\sqrt{\frac{8\pi G \rho_c}{3}}\bigg(\sum_i \Omega_{i,0}^{1-1/\delta_N}\rho_i^{1/\delta_N}\bigg)^{\delta_N/2}
\end{equation}
At $\delta_N=2$, we obtain $\alpha_i$ in Eq.(\ref{AO}).  Crucially, this proposed solution does not match observation in the radiation dominant epoch.
\\\\
However, Eq.(\ref{HTi}) can also admit a generic solution due to its underlying physical attributes, for example collisionless particles, so that, with $\alpha_i =1$,
\begin{equation}
(H_i^2)^{2 -\delta} =\bigg(\frac{8 \pi G}{3} \rho_i\bigg)^{2 - \delta}.
\label{Hi}
\end{equation}
which, for $\delta = 3/2$, becomes
\begin{equation}
H_i=\sqrt{\frac{8 \pi G}{3} \rho_i}.
\label{Hi}
\end{equation}
Unlike Eq.(\ref{AO}), Eq.(\ref{Hi}), with $\alpha_{rad} = 1$, is shown by JL to be in good agreement with observations in the radiation dominant epoch,  which means that amongst particles of a radiation fluid, the Boltzmann-Gibbs entropy is an effective description and the FLRW solution for a single fluid remains valid. We assume that Eq.(\ref{Hi}) is also true for single fluid baryons and single fluid dark energy. When two (or more) different fluids are present within the same horizon we may need a more appropriate entropy to account for the potential correlations between elements from distinct fluids. Tsallis entropy with the modified equations Eq.(\ref{STC})-Eq.(\ref{HT}) and the choice $\alpha_i = 1,$ becomes the simplest approach to address these considerations.
\\\\
In Eq.(\ref{HT}), the total Hubble parameter for Tsallis-Certo cosmology \cite{tsallis} is then, for $\delta=3/2$ and $\alpha_{rad} = 1$,
\begin{equation}
H_{TC}=H_r+H_b+H_{de}= \sqrt{\frac{8 \pi G}{3}}\bigg(\sqrt{\rho_{r}} +\sqrt{\rho_{b}} +\sqrt{\rho_{de }}  \bigg),
\label{HHH0}
\end{equation}
In particular,
\begin{eqnarray}
H_{TC}^2 &=& \frac{8 \pi G}{3}\bigg(\sqrt{\rho_{r}} +\sqrt{\rho_{b}} +\sqrt{\rho_{de }}  \bigg)^2
\nonumber
\\
&=& \frac{8 \pi G}{3}\bigg(\rho + 2\sqrt{\rho_{r}}\sqrt{\rho_{b}} + 2\sqrt{\rho_{b}}\sqrt{\rho_{de }} + 2\sqrt{\rho_{de }}\sqrt{\rho_{r}})
  \bigg)
\label{HHH1}
\end{eqnarray}
where $\rho = \rho_{r} + \rho_{b} + \rho_{de}$.
\\\\
We have not found it necessary to include CDM. With the definitions above, the $\omega CDM$ Hubble parameter is
\begin{equation}
H^2_{\omega CDM} = \frac{8 \pi G}{3}\bigg(\rho + \rho_{\omega CDM}\bigg)
\end{equation}
where CDM is played by some unknown collisionless particles. From (\ref{HHH1}), $H_{TC}$ also gives an effective CDM density,
\begin{equation}
\rho^{eff}_{\omega CDM} = 2\sqrt{\rho_{r}}\sqrt{\rho_{b}} + 2\sqrt{\rho_{de }}(\sqrt{\rho_{b}} + \sqrt{\rho_{r})}.
\end{equation}
Within the formalism of Mukhanov \cite{mukhanov}, we shall pursue a comparision of these effective densities in Section 4 where we find numerical agreement with CDM estimations at radiation-matter equality.
\\\\
We need a little more formalism first. Continuing to ignore CDM, at the  present time $z=0$, $H_{TC,0} = H_0$ by definition. The critical density $\rho_c$ which defines our flat universe is related to the measured Hubble constant by $H_0^2=\frac{8\pi G}{3}\rho_c$. This shows  that the "measured" critical density $\rho_c\neq \rho_0 = \rho_{r, 0} + \rho_{b, 0} + \rho_{de, 0}$.
\\\\
We can rewrite Eq.(\ref{HHH1}) as
\begin{eqnarray}
H_{TC}^2 = \frac{8 \pi G}{3} \rho_c\bigg(\frac{\rho}{\rho_c} + 2\sqrt{\frac{\rho_{r}}{\rho_c}}\sqrt{\frac{\rho_{b}}{\rho_c}} + 2\sqrt{\frac{\rho_{b}}{\rho_c}}\sqrt{\frac{\rho_{de }}{\rho_c}} + 2\sqrt{\frac{\rho_{de }}{\rho_c}}\sqrt{\frac{\rho_{r}}{\rho_c}}
  \bigg)
\label{HHH12}
\end{eqnarray}
Flatness is defined by the big bracket on the RHS (for all component densities there are) going to unity for $z=0$. Equivalently
\begin{eqnarray}
H_{TC}^2&=&\frac{8\pi G}{3}\rho_c\bigg(\sqrt{\frac{\rho_r}{\rho_c}} +\sqrt{\frac{\rho_b}{\rho_c}}+\sqrt{\frac{\rho_{de}}{\rho_c}}\bigg)^2
\nonumber
\\
&=&\frac{8\pi G}{3}\rho_c\bigg(\sqrt{\Omega_r^0}(1+z)^2+\sqrt{\Omega_b^0}(1+z)^{1.5} +\sqrt{\Omega_{de}^0 }(1+z)^{1.5(1+\omega}\bigg)^2,
\label{HH00}
\end{eqnarray}
where
\begin{equation}
\Omega_r^0 =\frac{\rho_{r,0}}{\rho_c}; \:\:\:\Omega_b^0=\frac{\rho_{b,0}}{\rho_c}.
\end{equation}
At $z=0$, we have the Hubble constant squared $H^2_{TC,0} =\frac{8 \pi G}{3}\rho_c$.
\begin{equation}
H_{TC}^2 =H_{TC,0}^2\bigg(\sqrt{\Omega_r^0}+\sqrt{\Omega_b^0}+\sqrt{\Omega_{de}^0}\bigg)^2.
\label{HH111}
\end{equation}
The flatness constraint then takes the form
\begin{equation}
\sqrt{\Omega_r^0}+\sqrt{\Omega_b^0}+\sqrt{\Omega_{de}^0}=1.
\end{equation}
The additivity in Eq.(\ref{HHH0}), written as 
\begin{equation}
H_{TC}\,r=H_r\,r+H_br +H_{de}\,r.
\end{equation}
is identical to adding free fall velocities.
This additivity condition is equivalent to the statement that the radiation FLRW frame, baryon FLRW frame and the dark energy FLRW frame are relatively non-comoving.  Conceptually speaking, this exhibits a {\it hidden} violation of the cosmological principle (CP) which could be important to the study of bulk flow discrepancy \cite{watkins} and galaxy-CMB dipole tension \cite{maartens1}-\cite{bohme}.
\\\\
Therefore, this relative non-comoving result between radiation frame, baryon frame (and dark energy frame) is the effect of the Tsallis-Cirto parameter $\delta$, which is the result of changing statistics within the apparent Horizon since, for $\delta =1$, we recover the standard Friedmann equation.
\\\\
The additivity of velocities from different frames can arise from the geodesic equations alone without explicitly identifying degrees of freedom.
In an earlier paper by us \cite{wong} we showed how an interpolating metric between Schwarzshild and Friedman-Robinson-Walker metrics (unique within its class) was such that a particle's effective free fall velocity is given by adding the free fall velocities due to the central mass and the expanding background. Recent work by de Haas \cite{deH} has shown that the same addition
equation follows from a biquartionic algebra that generates a relativistic
structure in the language of Qg rotors.
Indeed, this additivity of free falling velocities is a well established paradigm within Newtonian perturbation theory as proposed by Peebles \cite{peebles}.
\section{The Hubble parameter }
We need a better understanding of the  Hubble parameter $H_{TC}(z)$ in Eq.(\ref{HH00}), where the dominant components are radiation, baryons and dark energy. Although theoretically viable, it is imperative to see how far this $H_{TC}(z)$ can match observations. Given the severity of its formal difference to the FLRW formulation, it is not obvious that the matching is possible. Before we proceed we recall where the current Hubble parameter observations and modelling stand for conventional Friedmann equations ($\delta = 1$).
\subsection{$H(z)$ in the flat $\omega CDM$ model ($\delta = 1$)}
Since the observed Hubble parameter in DESI DR2 \cite{karim} is based on the $\omega CDM$ model, to obtain the observed Hubble parameter in our model, we borrow the formalism in \cite{karim}.
\\\\
The flat $\omega CDM$ model at redshift $z$, has a Hubble parameter
\begin{equation}
H_{\omega CDM} (z)=H_{0} \bigg( \Omega^0_{r} (1+z)^4 +\Omega^0_{\nu } (1+z)^{\alpha}+ \Omega^0_{c} (1+z)^3+\Omega_{b} (1+z)^3 +\Omega^0_{de } (1+z)^{3(1+\omega)}\bigg)^{1/2},
\label{H1}
\end{equation}
in which density parameters for cold dark matter $\Omega_{c}$ and neutrinos $\Omega_{\nu }$ are now included. The neutrino switches from radiation to dust at $1+z\sim m_{\nu}/(5\times 10^{-4}eV) $. We follow the CPL parametrisation used in \cite{karim} that $\omega=\omega_0+\omega_a(1-a)$ with $\omega_0$, $\omega_a$ being constant and $a$ is the scale factor (see Fig.1 from \cite{pf01}).  It is sufficient to say that, at larger values of $z$, the DE fluid has a phantom equation of state with $\omega<-1$, with negative kinetic energy. Efforts to describe this dark energy data with well known scalar field models are met with difficulties \cite{pf02}-\cite{pf03}. $\Omega_{r} h^2$, $\Omega_{b}h^2$ are also constrained by observations outside the CMB data. The CMB data in $\Lambda CDM$ model provides clear constraints on $\Omega_{c} h^2$ and $H_0$.
\\\\
There are increasing $H(z)$ observation data available for $z< 2.5$. For our purpose, we limit our work to comparing our model $H(z)$ with the Cosmic Chronometers (CC) data from Table 1 of \cite{jia}, which is based on a model independent approach proposed in \cite{jimenez}, and the DESI DR2 data \cite{karim}, which is a reliable data source for low redshifts upto to an effective redshift $z_{eff}=2.33$.
\\\\
To obtain $H(z)$ from the DESI DR2 data where the recombination redshift is taken at $z=1060$, we use the definition
\begin{equation}
H(z)=\frac{c}{D_H(z)} =\frac{r_d}{D_H(z)} \bigg(\frac{c}{r_d}\bigg),
\end{equation}
where $r_d$ is the sound horizon at recombination, $D_H(z)/r_d$ is an observed angular value given in Table IV in \cite{karim}. Although the CMB acoustic angular scale is an observed value, $r_d$ in our model may not be the same as that from $\Lambda CDM$.
\\\\
To illustrate this point, we recall that the CMB acoustic angular scale is given by
\begin{equation}
\theta_*=\frac{r_d}{D_M(z)}, \:\:\: r_d=\int_{1060}^{\infty} \frac{c_s(z)}{H(z)} dz,\:\:\:D_M(z=1060) =\int_0^{1060} \frac{cdz}{H(z)}.
\end{equation}
$D_M(z=1060)$ is the comoving angular diameter distance. To obtain the comoving sound horizon $r_d$ in the $\Lambda CDM$ model, we use
\begin{equation}
H_{\Lambda CDM} (z)=H_{0}\bigg(\Omega_{r} (1+z)^4+\Omega_{m} (1+z)^3+(1-\Omega_{m}-\Omega_{r}) \bigg)^{1/2}, \:\:\:
\end{equation}
with sound speed $c_s(z)$ at $z$
\begin{equation}
c_s^2(z)=\frac{c^2}{3(1+\frac{3}{4} \frac{\Omega_{b}}{\Omega_{r} (1+z)})}
\end{equation}
We use $H_0=70km/s/Mpc$, $\Omega_{m}=0.31$ for total matter and $\Omega_{r } h^2=2.47\times 10^{-5}$, $\Omega_{\nu }h^2=0.69\Omega_{r } h^2$ for relativistic neutrinos.
We find numerically that
\begin{equation}
r_d=\int_{1060}^{\infty} \frac{c_s(z)}{H^{\Lambda CDM} (z)}dz =\frac{c}{H_0} 0.03377=144.68 Mpc.
\end{equation}
The reported value in Planck is $r_d=144.43\pm 26 Mpc$ \cite{planck}. For $D_M(z)$ we keep the neutrino relativistic (which could incur an error $\lsim 1\%$) such that
\begin{equation}
D_M(z=1060)= \int_0^{1060} \frac{cdz}{H^{\Lambda CDM}(z)}=\frac{c}{H_0}3.138.
\end{equation}
This gives a model CMB angular acoustic scale at
\begin{equation}
100\theta_*=1.07,
\end{equation}
where the observed CMB value at $100\theta_*=1.04$. This observed CMB angular scale plays the role of a critical constraint for filtering the many extended models of $\Lambda CDM$.
\subsection{$H_{TC}(z)$ in Tsallis-Cirto Cosmology ($\delta = 3/2$)}
We can now perform similar calculations using our model Hubble parameter taking into account of the behaviour of neutrinos. We note from \cite{karim} that the neutrino mass $m_{\nu}<0.042eV$ and therefore its equation of state changes from radiation to pressureless matter at $1+z\sim m_{\nu}/(5\times 10^{-4}eV) \sim 100$, which does not generate a sustained $H_{\nu}$ over a large period of time. Before $z\sim100$, neutrinos are considered relativistic and we can treat its density as radiation density as before. We assume that the Hubble parameter takes the following form for $100 \leq z<\infty$, where $\Omega_{r \nu }=\Omega_{r}+\Omega_{\nu }$.
\\\\
We have
\begin{equation}
H_{TC,bf}(z) =H_{0} \bigg( \sqrt{\Omega^0_{r \nu }}(1+z)^2+ \sqrt{\Omega^0_{b }} (1+z)^{1.5} +(1-\sqrt{\Omega^0_{r \nu }}-\sqrt{\Omega^0_{b }})(1+z)^{1.5(1+\omega)}\bigg)
\label{Hz1}
\end{equation}
Based on the relation $\rho_{\nu }= N_{eff} (7/8)(4/11)^{4/3}\rho_{r}$, with number of neutrino generations $N_{eff}=3.046$ \cite{planck}, we take $\Omega_{r \nu }h^2=1.69\Omega_{r }h^2=1.69\times 2.47\times 10^{-5}$.
\\\\
After $z=100$, we identify neutrino as dust and take the Hubble parameter, where $\Omega_{b \nu }=\Omega_{b }+\Omega_{\nu }$, as
 \begin{equation}
H_{TC,af}(z) =H_{0} \bigg( \sqrt{\Omega^0_{r} }(1+z)^2+ \sqrt{\Omega^0_{b \nu }} (1+z)^{1.5} +(1-\sqrt{\Omega^0_{r}}-\sqrt{\Omega^0_{b \nu }})(1+z)^{1.5(1+\omega)}\bigg).
\label{Hz2}
\end{equation}
Here we use $\Omega^0_{\nu}=0.001$.
\\\\
Considering the cosmological parameters in the RHS of Eq.(\ref{Hz1})-Eq.(\ref{Hz2}), the $\Omega_r h^2$ and $\Omega_{\nu} h^2$ can be treated as model independent. In the past, BBN yields an estimate for $\Omega_bh^2=0.020 \pm 0.0015$ \cite{burles}. This value has been updated to $\Omega_b h^2=0.0223\pm 0.00055$ \cite{schoneberg}, which is almost identical to the Planck CMB value \cite{planck}. We therefore consider only the range $0.02175\leq \Omega_bh^2 \leq0.02285$. Given the current debate on the observed Hubble constant value \cite{freedman}-\cite{divalentino}, we consider Hubble constant in the range $h=0.7-0.74$. {For convenience, we work with $c_0+c_a(1-a)=\frac{3}{2}(1+\omega) =1.5(1+\omega_0+\omega_a(1-a))$.
\\\\
The small observed CMB angular acoustic scale $100\theta_*=1.04$ imposes a non-trivial constraint on our model parameters.
Our primary source of $H(z)$ data in this work is from the model independent CC. For $z\leq 1$, the dark energy term $\sqrt{\Omega_{de}} (1+z)^{c_0+c_a(1-a)}$ is not far from a constant dark energy. By varying $c_0<0.25$ and $c_a\leq1$, we find that the H(z) can fit the data well. For CC data between $1<z<2$, we find by trials the minimal values $c_0=0.1$, $c_a=0.85$ can fit the CC data very well.
\\\\
We next calculate the angular acoustic scale $\theta_*=r_d/D_M$ and find that
for $h=0.70$, $\Omega_b h^2=0.02175$, $c_0=0.2$, $c_a=0.85$ such that $\frac{3}{2}(1+\omega)=0.2+0.85(1-a)$, one obtains
\begin{equation}
r_d = \int_{1060}^{\infty} \frac{c_s(z)}{H_{bf}^{TC} (z)}dz=\frac{c}{H_0} 0.0374.\:\:\:
\label{rdtc}
\end{equation}
\begin{equation}
 D_M(z=1060) = \int_{100}^{1060} \frac{cdz}{H_{bf}^{TC}(z)}+\int_0^{100} \frac{dz}{H_{af}^{TC}(z)}=\frac{c}{H_0}3.606.
\end{equation}
such that the model angular scale at recombination is
\begin{equation}
100\theta_*=1.037.
\end{equation}
Since DESI DR2 only provide data for $r_d/D_H(z)$,  the "observed" $H(z)$ within our model TC cosmology is then obtained using Eq,(\ref{rdtc})
\begin{equation}
H(z)=\frac{c}{D_H(z)} =\bigg(\frac{r_d}{D_H(z)}\bigg) \frac{H_0}{0.0374}.
\label{Hzdr2}
\end{equation}
We note that the error bar of the observed $D_M(z)/r_d$ is in general small ($\leq 2\%$), but there are small flexibilities in our combined choice of $H_0$, $c_0$ and $c_a$, so that for our purpose, we allow for a error bar of $\sim 5\%$ for the derived DESI $H(z)$ from Eq.(\ref{Hzdr2}).
\\\\
For a fixed set of $c_0, c_a$, the angular scale $r_d/D_M$ is independent of $H_0$. We extend the parameter set to $c_0=0.2$ and $c_a=1.0$ which turns out to be parameters' upper limit when the SMBH growth rate is taken into consideration, this is given elsewhere.
\begin{figure}[p]
\begin{center}
\resizebox{150mm}{!}{\includegraphics{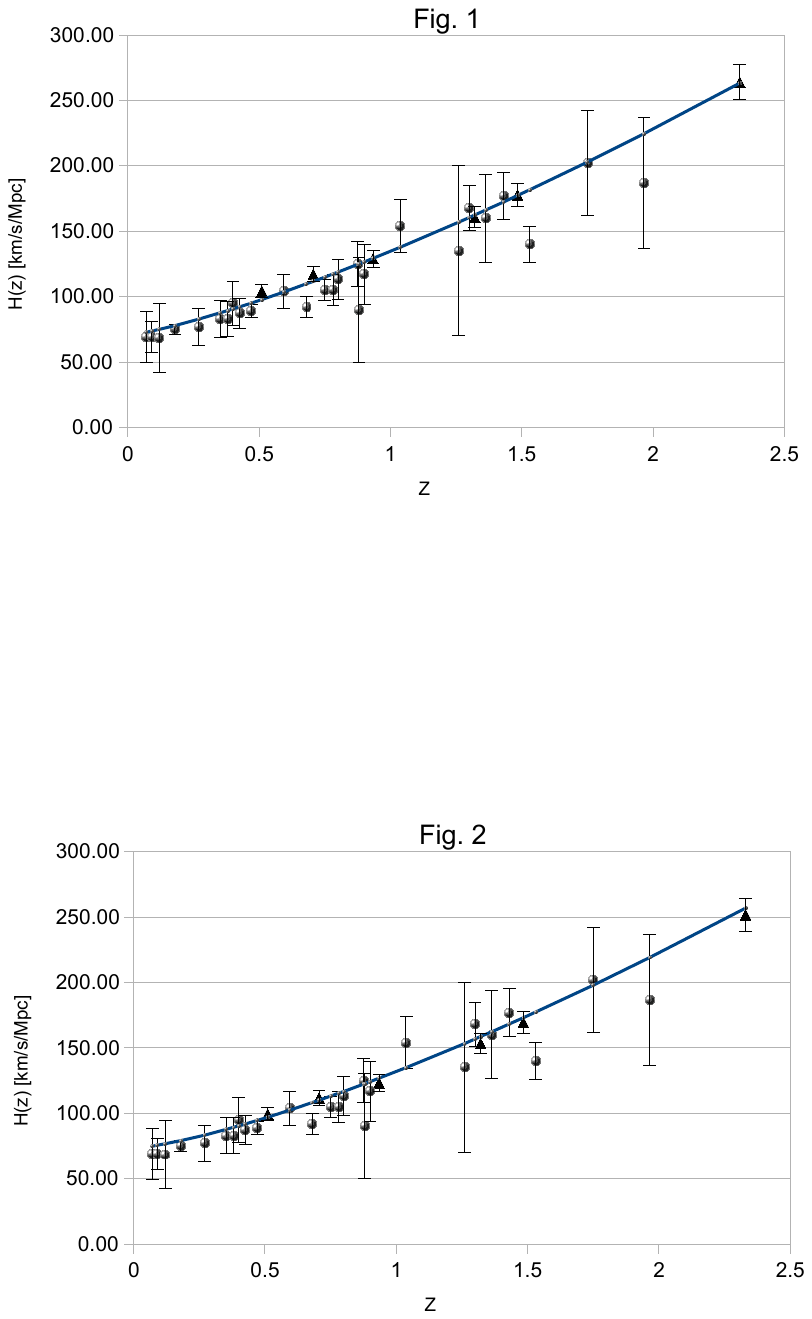}}
\end{center}
{\footnotesize Fig. 1: Plot of model H(z) (solid line) against the CC data (Dots) and DESI DE2 data (Triangles). The cosmological parameters are $h=0.70$, $\Omega_bh^2=0.02175$, $\omega_0=-0.8466$, $\omega_a=0.666$.}
\\\
{\footnotesize Fig. 2: Plot of model H(z) (solid line) against the CC data (Dots) and DESI DE2 data (Triangles). The cosmological parameters are $h=0.73$, $\Omega_bh^2=0.02175$, $\omega_0=-1$, $\omega_a=0.83$.}
\end{figure}

\subsubsection*{Plotting the data}
For $h=0.70$, we plot our model $H(z)$ (which is $H_{TC}(z)$) vs the CC data and the model consistent observed $H(z)$ data derived from DESI DR2,  the result is shown in Fig.1. We see that our model $H(z)$ values match the DESI data and the $z<1$ CC data very closely. The large error bar CC data at $z>1$ also fit well to our model $H(z)$. We repeat the same process for higher $h$ values at fixed  $\Omega_bh^2 =0.02175$ and obtain the following,
\\\\
For $h=0.71$, we find $\omega_0=-0.866,\: \omega_a =0.666$, $\frac{3}{2}(1+\omega)=0.2+(1-a)$ with $100\theta_*=104.2$.
\\\\
For $h=0.72$, we find $\omega_0=-0.873,\:\omega_a=0.666$, $\frac{3}{2}(1+\omega)=0.19+(1-a)$ with $100\theta_*=104.7$.
\\\\
For $h=0.73$, we find $\omega_0=-1, \: \omega_a=0.82$, $\frac{3}{2}(1+\omega)=1.23(1-a)$ with
$100\theta_* =103.6$.
\\\\
For $h=0.74$, we find $\omega_0=-1,\: \omega_a=0.813$, $\frac{3}{2}(1+\omega)=1.22(1-a)$ with $100\theta_8=104.2$.
\\\\
\newpage
\noindent We note that for a narrow range $0.15<c_0<0.2$ and $0.85<c_a<1$, the plot of $H(z)$ will fit CC and DESI data as well as keeping $r_d/D_M$ close to $\theta_*=1.04$.
The resulting $\omega$ is alway $\omega>-1$ and therefore shows no phantom crossing. The plots of $H(z)$ for $h=0.71, 0.72, 0.73$,  fit data similarly well as Fig.1. We present the plot for $h=0.73$ with its corresponding choice of CPL parameters in Fig.2. We note that $h=0.735\pm 0.0081$ is the most accurate result from distance ladder approach to date \cite{riess3} .
\\\\
We can estimate the deceleration-acceleration equality redshift in this model. For $h=0.72$,
\begin{equation}
\omega=\omega_{de}=-0.207-0.666a.
\end{equation}
The late time energy density becomes
\begin{equation}
\rho =\rho_b+2\sqrt{\rho_b\rho_{de} }+\rho_{de},
\end{equation}
with the acceleration equation for these fluids is
\begin{eqnarray}
&&\frac{\ddot{a}}{a}=-\frac{4\pi G}{3} \sum_i \rho_i(1+3\omega_i) =
\\
&& = -\frac{4\pi G\rho_{c0} }{3} (1+z)^3\bigg(\Omega_b+
+\Omega_{de} a^{(0.62+2a)} (1+3\omega_{de})
+2\sqrt{\Omega_b\Omega_{de}} a^{(1.81+a)}(1+\frac{3}{2}\omega_{de})\bigg)
\nonumber
\label{acc}
\end{eqnarray}
Consider the terms in the big bracket in Eq.(\ref{acc})
\begin{equation}
\frac{\ddot{a}}{a} \propto -\bigg(0.044+0.624a^{0.62+2a}(0.379-2a) +0.324a^{1.81+a}(0.688-a)\bigg).
\end{equation}
For $a<0.189$ ($z>4.27$), every term in the RHS bracket is positive and we have deceleration $\ddot{a}/a<0 $, and for $a=1$ the bracket on the RHS is negative and we have acceleration $\ddot{a}/a>0$.
We start with a value of the scale factor $a$ in the range $1<z<2$ and find recursively that $\ddot{a}/a\sim 0$ slowly across
\begin{equation}
a\sim 0.37,\:\:z\sim1.7.
\end{equation}
Our solution shows that a consistent TC cosmology without CDM is possible.
\section{A Note on the Constraint from the Acoustic Peak Height}
Finally, we see that this cosmology can reproduce
the CMB angular peak heights without postulating CDM.
\\\\
We repeat Eq.(\ref{HHH1}),
\begin{equation}
H_{TC}^2=\frac{8\pi G}{3} \bigg(\rho_r+\rho_b+\rho_{de} +2\sqrt{\rho_r\rho_b}+2\sqrt{\rho_r\rho_{de}}+2\sqrt{\rho_b\rho_{de}}\bigg)
\label{HHH1a}
\end{equation}
 In general, dark energy (constant or dynamical) does not associate with a growing density perturbation \cite{mukhanov}, therefore we only consider the term $\delta_{rb} \rho_{rb}$ as the potential candidate for dark matter potential.
In Eq.(\ref{HHH1a}), we see that adding the radiation and baryon Hubble parameters introduces an effective energy density $\rho_{rb}= 2\sqrt{\rho_r \rho_b}$ in the Friedmann equation, the radiation-matter equality occurs at $\rho_r=2\sqrt{\rho_r \rho_b}+\rho_b$ with solution
\begin{equation}
\rho_r=(1+\sqrt{2})^2\rho_b=5.828\rho_b,
\label{rhorrhob}
\end{equation}
where the non-baryonic matter density at equality is $4.828 \rho_b$, while the equality CDM density takes the value
\begin{equation}
\rho_r=5.826\rho_b,
\label{rb}
\end{equation}
required in WMAP \cite{liddle}. This leads to the same equality redshift $z_{eq}$ as the $\Lambda CDM$ model where, by dividing Eq.(\ref{rb}) by the critical denesity $\rho_c$, one obtains $\Omega_{r,0} (1+z_{eq})=5.826\Omega_{b,0}$.
\\\\
In modelling for CMB angular power spectrum, the dark matter density parameter $\Omega_{c} h^2$ is further constrained by the acoustic peak heights via its gravitational potential in conformal time $\eta$ \cite{mukhanov},
\begin{equation}
\phi_k(\eta>\eta_{eq})  \sim - \frac{4\pi Ga^2}{k^2} \rho_{c} \delta_{c} |_{\eta_{eq}},
\label{phik1}
\end{equation}
where $\delta_c$ is the dark matter overdensity and $\eta_{eq}$ is the conformal time at radiation-matter equality. For CDM, $\rho_c\propto a^{-3}$ and $\delta_c \propto a$, one can see that this potential term is constant for fixed $k$.   The conformal time is given as
\begin{equation}
\eta=\int_0^{t} \frac{1}{a} dt =-\int_z^{\infty} \frac{dz}{H(z)}
\end{equation}
The integration range is mainly in the radiation dominant epoch, $\eta_{eq}$ is similar for $H^{\Lambda CDM}(z)$ and $H^{TC}(z)$.
$\rho_{rb} =2\sqrt{\rho_r \rho_b}\propto a^{-7/2}$ and we find that $\delta_{rb}=\delta\rho_{rb}/\rho_{rb}=\sqrt{\delta_r\delta_b} \propto \sqrt{a^2 a}=a^{3/2}$ so that the underlying non-baryonic matter potential takes the form,
\begin{equation}
\phi_k(\eta>\eta_{eq}) \sim -\frac{4\pi Ga^2}{k^2} \rho_{rb} \delta_{rb} |_{\eta_{eq}}\propto - \frac{4\pi Ga^2}{k^2} \rho_{c} \delta_{c} |_{\eta_{eq}}
\label{phik}
\end{equation}
One can see that both potentials are constant for fixed $k$. At equality, for adiabatic expansion where $\delta S$=0, we have
\begin{equation}
\frac{\delta S}{S} =\frac{3}{4}\delta_r-\delta_b=0,
\end{equation}
\begin{equation}
\delta_{rb}=\sqrt{\delta_r \delta_b}=\sqrt{\frac{4}{3}}\delta_b=\sqrt{\frac{4}{3}}\delta_c,
\end{equation}
since dark matter overdensity only grows logarithmically at radiation dominant epoch, at radiation-matter equality it is usually taken as $\delta_c=\delta_b$ \cite{mukhanov}. Using Eq.(\ref{rhorrhob}) where $\Omega_{r} (1+z_{eq})=5.828\Omega_{b}$, so that at equality redshift we obtain
\begin{equation}
\rho_{rb}\delta_{rb} =\rho_{cr}\bigg(2\sqrt{ \Omega_{r} \Omega_{b} (1+z_{eq})} (1+z_{eq})^3\bigg) \sqrt{\frac{4}{3}}\delta_c=5.576\Omega_{b} \rho_{cr} (1+z_{eq})^3\delta_c=\rho_c\delta_c.
\end{equation}
In this epoch, this non-baryonic potential $\phi_k(\eta>\eta_{eq})$ can be considered as an "effective" dark matter potential with $\Omega_{c}h^2 =5.576 \Omega_{b} h^2$. For $\Omega_{b} h^2=0.02175$, we obtain $\Omega_{c} h^2= 0.1212$ which is about two percent at variant with the Planck value $\Omega_{c} h^2=0.1186\pm 0.0020$ \cite{planck}.
\\\\
This means that the $\Lambda CDM$ constraint on dark matter density parameter coming from the observed CMB acoustic peak heights could be satisfied by the TC model without any fine-tuning. In other words, this TC model does not impose a strong constraint on the Hubble constant value as would be the case if the underlying model is $\Lambda CDM$ and therefore is likely to evade the Hubble tension argument.
\section{ Summary and Conclusion}
Tsallis-Cirto (TC) entropic cosmology \cite{tsallis} characterises the effect of the interactions between system elements of the universe through a parameter $\delta$. When $\delta=1$, the entropy of a simple (collisionless and non-interacting) fluid in 3-dimensions is determined by the surface area A of its horizon such that the entropy is extensive and additive. The first law of thermodynamics with this entropy leads to the standard Friedmann equation. For a system with multiple fluid subsystems, if there are no additional correlations between elements from distinctly different fluids, we arrive at the standard model of cosmology $H^2=\frac{8 \pi G}{3}\sum_i \rho_i$.
\\\\
However, if  correlations between elements from distinctly different fluids exist, one could attempt to capture their effects on entropy using Tsallis entropy at some $\delta>1$. At $\delta=\frac{3}{2}$ where the entropy depends extensively on its physical system size (volume), we find that the Hubble parameter of the composite system is again separable into its subsystem  Hubble parameters. Jizba and Lambiase \cite{jizba} show that in the radiation dominant epoch, the standard Friedmann equation for radiation only universe $H_r=\sqrt{\frac{8 \pi G}{3}\rho_r}$ provides  good agreement with BBN and primordial light element abundance data for $\delta = 3/2$. Assuming that this Friedmann equation is true for a baryon and dark energy only universe, we find a solution in which the Hubble parameter is obtained by adding the individual asymptotic Hubble parameters for radiation, baryon and dark energy.
\\\\
This result means that  individually a cosmic background fluid follows its own FLRW frame, but they are relatively non-comoving and do not share a single collective FLRW frame. In this sense, the Cosmological Principle is modified.
\\\\
We then investigate, using the CPL parametration of dark energy, how far this TC Hubble parameter at lower redshifts can match data from  Cosmic Chronometers (CC) and DESI DR2. The bulk of data comes from CC, which is model independent. To account for CC data at  $0<z<2$ whilst keeping $100\theta_*=1.04$ constraint, we parametrise dark energy to match CC data. We then obtain $H(z)$ using this parametrisation with DESI DR2 data and find that these data are consistent with the CC data. 
\\\\
This ensures that the dark energy equation of state $\omega >-1$, and therefore remains non-phantom at late times.
Finally, we see that Tsallis-Cirto ($\delta = 3/2$) cosmology can reproduce
the CMB angular peak heights without postulating CDM.
%

%
%\begin{figure}[p]
%\begin{center}
%\resizebox{150mm}{!}{\includegraphics{Newplot2025.pdf}}
%\end{center}
%Fig. 1: Plot of model H(z) (solid line) against the CC data (Dots) and DESI DE2 data (Triangles). The cosmological parameters are $h=0.70$, $\Omega_bh^2=0.02175$, $\omega_0=-0.8466$, $\omega_a=0.666$.
%\\\
%Fig. 2: Plot of model H(z) (solid line) against the CC data (Dots) and DESI DE2 data (Triangles). The cosmological parameters are $h=0.73$, $\Omega_bh^2=0.02175$, $\omega_0=-1$, $\omega_a=0.83$.
%\end{figure}

\begin{thebibliography}{99}
%
\bibitem{karim} Abdul Karim, M., et al.  DESI DR2 Results II: Measurements of Baryon Acoustic Oscillations and Cosmological Constraints,
https://doi.org/10.48550/arXiv.2503.14738
%
\bibitem{ozulker} $\ddot{O}$z$\ddot{u}$lker, E., Di Valentino, E., Giar$\acute{e}$, W.  Dark Energy Crosses the Line: Quantifying and Testing the Evidence for Phantom Crossing,
 https://doi.org/10.48550/arXiv.2506.19053
%
\bibitem{scherer} Scherer, M., Sabogal, M., Nunes, R. C., De Felice, A. Challenging $\Lambda CDM$: 5$\sigma$ Evidence for a Dynamical Dark Energy Late-Time Transition,
https://doi.org/10.48550/arXiv.2504.20664
%
\bibitem{caldwell} Caldwell, Robert R.; Kamionkowski, Marc; Weinberg, Nevin N. (2003). "Phantom Energy and Cosmic Doomsday". Physical Review Letters. 91 (7) 071301. arXiv:astro-ph/0302506. Bibcode:2003PhRvL..91g1301C. doi:10.1103/PhysRevLett.91.071301. PMID 12935004. S2CID 119498512.
%
\bibitem{tsallis} Tsallis, C., Cirto, L. J. L  Eur. Phys. J. C 73, 2487 (2013). arxiv.org/pdf/1410.8591, arxiv.org/pdf/1202.2154.
%
\bibitem{jizba0} Jizba, P., Lambiase, G. Tsallis cosmology and its applications in dark matter physics with focus on IceCube high-energy neutrino
data. Eur. Phys. J. C 2022, 82, 1123. https://doi.org/10.1140/epjc/s10052-022-11113-2
%
\bibitem{jizba}  Jizba P.,  Lambiase, G. Entropy 25, 1495 (2023), arxiv.org/pdf/2310.19045
%
\bibitem{tsallis2} Tsallis, C. Black Hole Entropy; A Closer Look. Entropy 2020, 22,17
%
\bibitem{tsallis3} Tsallis, C., Jensen, H. J.,   Extensive Composable Entropy for the Analysis of Cosmological Data, arxiv.org/pdf/2408.08820
%
\bibitem{jensen_jizba} Jensen, H.J., Jizba, P and Tempesta, P. Universality classes, Thermodynamics of Group Entropies, and Black Holes, arXiv:2603.02385v1
%
\bibitem{moha1} Khodam-Mohammadi, A., Monshizadeh, M.  Physics Letters B 843, 138066 (2023), ISSN
0370-2693, URL http://dx.doi.org/10.1016/j.physletb.2023.138066.
%
%
\bibitem{moha2} Khodam-Mohammad, A., Monshizadeh, M., Cosmological Tensions with Non-Extensive Entropic Cosmology:
A Modified Stress-Energy Approach,  arxiv.org/pdf/2509.19845.
%
\bibitem{beken1}  Bekenstein, J.D.  Phys. Rev. D 7, 2333 (1973)
%
\bibitem{beken2} Bekenstein, J. D. Phys. Rev. D 9, 3292 (1974)
%
\bibitem{hawking1} Hawking, S. K. Nature 248, 30 (1974)
%
\bibitem{hawking2}  Hawking, S. K. Phys. Rev. D 13, 191 (1976)
%
%
\bibitem{barrow} Barrow, J. D. arXiv:2004.09444 [gr-qc].)

%

\bibitem{jia} Jia, X. D., Hu,  J. P., Wang, F. Y.  Evidence of a decreasing trend for the Hubble constant, A and A 674, A45, 2023,
https://doi.org/10.1051/0004-6361/202346356, arXiv:2212.00238v3
%
\bibitem{krishnan} Krishnan, C., et al. Running Hubble tension and a H0 diagnostic,  Phys. Rev. D 103, 103509 (2021). https://doi.org/10.1103/PhysRevD.103.103509
%
\bibitem{loeb2} Chen, X., Loeb, A. Evolving Dark Energy or Evolving Dark Matter? https://doi.org/10.48550/arXiv.2505.02645
%
\bibitem{dwang} Wang, D., Bamba, K. ACT DR6 Leads to Stronger Evidence for Dynamical Dark Matter, https://doi.org/10.48550/arXiv.2506.23029

%
\bibitem{kroupa6} Samaras, N., Grandis, S.,  Kroupa, P. On the initial conditions of the $\nu$HDM cosmological model, accepted for publication in MNRAS,
https://doi.org/10.48550/arXiv.2506.19196
%
\bibitem{akarsu} Akarsu, O., Barrow, J. D., Escamilla, L. A., Vazquez, J. A., Phys. Rev. D 101, 063528 (2020). arXiv:1912.08751.
%
\bibitem{zimdahl} Zimdahl, W. Interacting Dark Energy and Cosmological Equations of State, International Journal of Modern Physics D, Vol14, No. 12, pp2310-2325 (2005)
%

%
\bibitem{lopez} Lopez-Corredoira, Marmet, M. L. Alternative ideas in cosmology, invited review to be published in Int. J. Mod. Phys. D 2202.12897,
https://doi.org/10.1142/S0218271822300142
%
\bibitem{jacobson}  Jacobson, T.  Phys. Rev. Lett. 75, 1260 (1995).
%
\bibitem{verlinde} Verlinde, E. JHEP 1104, 029 (2011).
%
\bibitem{pad} Padmanabhan, T. arXiv:1206.4916

\bibitem {sheykhi} Sheykhi, A. Phys. Lett. B 785, 118 (2018).
%
%
\bibitem{akbar} Akbar, M.,  Cai,  R. G. Phys. Rev. D 75, 084003 (2007).
%
%
\bibitem{mukhanov} Mukhanov, V. Physical foundation of cosmology, Cambridge University Press 2005
%
%\bibitem{navas} Navas, S. et al. Review of Particle Physics, PRD. 110(3). pp.030001 (2024).
%
%
\bibitem{watkins} Watkins, R., et al. Analyzing the Large-Scale Bulk Flow using CosmicFlows4: Increasing Tension with the Standard Cosmological Model,
https://doi.org/10.1093/mnras/stad1984
%
\bibitem{maartens1} Guandalin, C.,  Piat, J., Clarkson, C., Maartens, R. Theoretical systematics in testing the Cosmological Principle with the kinematic quasar dipole, Astrophys. J. 953 (2023) 2, 144,
https://doi.org/10.3847/1538-4357/acdf46

\bibitem{ellis2}
Ellis, G.,  Baldwin,  J. 1984, Monthly Notices of the Royal
Astronomical Society, 206, 377,
doi: 10.1093/mnras/206.2.377

%
\bibitem{secrest1} Secrest, N. J., von Hausegger, S., Rameez, M., Mohayaee,
R.,  Sarkar, S. 2022, The Astrophysical Journal Letters,
937, L31, doi: 10.3847/2041-8213/ac88c0
%
\bibitem{secrest2}
Secrest, N. J., von Hausegger, S., Rameez, M., et al. 2021,
The Astrophysical journal letters, 908, L51,
doi: 10.3847/2041-8213/abdd40
%
\bibitem{bohme} B$\ddot{o}$hme, L., et al. Overdispersed radio source counts and excess radio dipole detection, 	Phys. Rev. Lett. 135, 201001 (2025),
https://doi.org/10.1103/6z32-3zf4
%
\bibitem{wong}  Wong, C. C.,  Rivers, R. J. Variable Modified Newtonian Mechanics I: The Early Universe, arXiv: 1601.00376. https://doi.org/10.48550/arXiv.1601.00376
%
\bibitem{deH} de Haas, P., Constructing Spacetime from Rapidity: Explicit
BQ Derivation of the PG coframe and the metric tensor,
DOI:10.5281/zenodo.17717365
%
\bibitem{peebles} Peebles, P. J. E., The Large Scale Structure of the Universe, Princeton University Press 1980
%
%
%\bibitem{tsagas} Tsagas, C. G., Perivolaropoulos, L., Asvesta, K., Large-scale peculiar
%velocities in the universe, Invited review to appear in Phys. Rep.
%https://doi.org/10.48550/arXiv.2510.05340
%
\bibitem{pf01} Wolf, W. J.,  Ferreira, P. G. Underdetermination of dark energy, Phys. Rev. D 108 (2023) 103519,  https://doi.org/10.1103/PhysRevD.108.103519
%
\bibitem{pf02} Wolf, W. J.,  García-García, C., Bartlett, D. J., Ferreira, P. G. Scant evidence for thawing quintessence, Phys. Rev. D 110 (2024) 083528,
https://doi.org/10.1103/PhysRevD.110.083528
%
\bibitem{pf03} Wolf, W. J.,  García-García, C., Theodore, A., Ferreira, P. G. Assessing cosmological evidence for non-minimal coupling, 	Phys. Rev. Lett. 135 (2025) 081001,
https://doi.org/10.1103/jysf-k72m
%
%
\bibitem{jimenez} Jimenez, R., Loeb, A.   ApJ, 573, 37 (2002).
https://doi.org/10.1086/340549
%
%
\bibitem{planck} Planck Collaboration VI, Planck 2018 results. VI. Cosmological parameters, A and A 641, A6 (2020), https://doi.org/10.1051/0004-6361/201833910
%

\bibitem{burles} Burles, S.,  Nollett, K., Turner, M. 2000, Astrophys.J. 552 (2001) L1-L6.
https://doi.org/10.1086/320251
%
\bibitem{schoneberg} Schoneberg, N. The 2024 BBN baryon abundance update,
https://doi.org/10.48550/arXiv.2401.15054
%
\bibitem{freedman} Freedman, W. L., et al.  Status Report on the Chicago-Carnegie Hubble Program (CCHP): Measurement of the Hubble Constant Using the Hubble and James Webb Space Telescopes, https://doi.org/10.48550/arXiv.2408.06153

%
\bibitem{divalentino} Di Valentino, E., et al., The CosmoVerse White Paper: Addressing observational tensions in cosmology with systematics and fundamental physics, accepted in PotDU, https://doi.org/10.48550/arXiv.2504.01669
%
%
\bibitem{riess3} H0DN Collaboration, The Local Distance Network: a community consensus report on the measurement of the Hubble constant at 1% precision, Astronomy and Astrophysics , 708, A166 (2026),
https://doi.org/10.48550/arXiv.2510.23823,
%

\bibitem{liddle} Lahav, O.,  Liddle, A. R.  The cosmological parameters in The Review of particle physics, Eidelman, S. et al. (Particle Physics Group), Phys. Lett. B592, 1 (2004),
https://doi.org/10.48550/arXiv.astro-ph/0406681

\end{thebibliography}
\end{document}